\documentclass[prl,twocolumn,noshowpacs,preprintnumbers,amsmath,amssymb,longbibliography]{revtex4-1}

\usepackage{amsfonts,amsthm,amsmath,amssymb,mathrsfs,mathtools}
\usepackage{dsfont}
\usepackage{dcolumn}
\usepackage{epsfig, color} 
\usepackage{verbatim,hyperref,graphics,graphicx,textcomp}
\usepackage{epstopdf}

\newcommand*{\diff}{\mathop{}\!\mathrm{d}}
\newcommand{\derpar}[2]{\frac{\partial #1}{\partial #2}}
\newcommand{\dersecpar}[2]{\frac{\partial^2 #1}{\partial #2^2}}

\newcommand{\curl}[1]{\nabla\times #1}
\newcommand{\vecomega}{\boldsymbol{\omega}}
\newcommand{\vecxi}{\boldsymbol{\Xi}}
\renewcommand{\vec}{\mathbf}
\renewcommand{\Re}{\mathrm{Re}\,}
\newcommand{\mrI}{\mathbb{I}}

\newcommand{\mrD}{\mathbb{D}}

\newcommand{\SM}{\cite{CairoliF}}

\newcommand{\beq}{\begin{equation}}
\newcommand{\eeq}{\end{equation}}
\newcommand{\beqn}{\begin{eqnarray}}
\newcommand{\eeqn}{\end{eqnarray}}

\begin{document}

\title{Hydrodynamics of Active L{\'e}vy Matter}

\author{Andrea Cairoli}\email{andrea.cairoli@crick.ac.uk}
\altaffiliation[current author's affiliation: ]{The Francis Crick Institute, London NW1 1AT, United Kingdom.}
\affiliation{Department of Bioengineering, Imperial College London, London SW7 2AZ, United Kingdom}
\author{Chiu Fan Lee}\email{c.lee@imperial.ac.uk}
\affiliation{Department of Bioengineering, Imperial College London, London SW7 2AZ, United Kingdom}

\begin{abstract}

Collective motion 
is often modeled within the framework of active fluids,
where the constituent active particles, when interactions with other particles are switched off, 
perform normal diffusion at long times. 
However, in biology, single-particle superdiffusion and fat-tailed displacement statistics are also widespread. 
The collective properties of interacting systems exhibiting such anomalous diffusive dynamics, 
which we call \textit{active L{\'e}vy matter}, 
cannot be captured by current active fluid theories. 
Here, we formulate a hydrodynamic theory of active L{\'e}vy matter 
by coarse-graining a microscopic model of 
aligning polar active particles that perform superdiffusion akin to L{\'e}vy flights. 
Applying a linear stability analysis on the hydrodynamic equations at the onset of collective motion, we
find that, in contrast to its conventional counterpart, the order-disorder transition can become critical. 
We then estimate 
the corresponding critical exponents 
by finite size scaling analysis of numerical simulations.
Our work highlights the novel physics in active matter that integrates both anomalous diffusive motility and inter-particle interactions.

\end{abstract}

\maketitle

Active matter refers to systems comprising entities with 
the ability to generate directed motion perpetually
\cite{Toner2005,Schweitzer2007,Ramaswamy2010,Marchetti2013,Hauser2015,Needleman2017}.  
Interactions among these constituent units can
cause the spontaneous emergence of collective behavior that includes 
collective motion 
\cite{Vicsek1995,Toner1995,Toner1998,Vicsek2012},
turbulent patterns 
\cite{Dombrowski2004,Hernandez-Ortiz2005,Sokolov2007,Aranson2007,Saintillan2007,Wolgemuth2008,
Sanchez2012,Wensink2012,Doostmohammadi2018},
and motility-induced phase separation 
\cite{Tailleur2008,Fily2012,Redner2013,Cates2015}.
To study a many-body systems of this kind 
a key objective is to derive a set of equations 
that can capture the coarse-grained behavior of the system from the corresponding microscopic dynamics.
A simple and yet successful method to achieve this task is to first decide on a set of hydrodynamic variables,
and then write down the most general possible set of equations of motion (EOM) based on an expansion of these variables and their spatial derivatives, 
where the only constraints are the conservation laws and symmetries of the underlying microscopic dynamics. 
Two classic examples are the hydrodynamic theories of thermal fluids and momentum non-conserving active fluids, where the hydrodynamics variables are the density and velocity fields, whose dynamics are described by the Navier-Stokes \cite{Chaikin2000} 
and Toner-Tu equations \cite{Toner1995,Toner1998}, respectively. 
To incorporate fluctuations into the EOM, Gaussian noise terms are typically added, 
which are justified by the central limit theorem as it reflects the fact that averaging over many microscopic fluctuations with finite variances (temporally and/or spatially) inevitably leads to a Gaussian distribution \cite{Gardiner2009}.


However, many natural and social phenomena can also exhibit fluctuations that are well approximated by random variables drawn from distributions with divergent variances, e.g., power-law tailed distributions. 
In the  biological context 
these fluctuations can manifest as the anomalous superdiffusive dynamics of the constituent units \cite{Metzler2000,Cavagna2013,Murakami2015,Ariel2015}
(which may constitute 
an optimal foraging strategy for living organisms 
\cite{Viswanathan1999,Lomholt2008levy,
Benichou2011,Viswanathan2011}),
or can emerge from nonlinear and/or memory effects \cite{Fedotov2017}.
Remarkably, the average of these random variables 
can again converge to a universal distribution, called the $\alpha$-stable L\'evy distribution, 
as guaranteed by the {\it generalized} central limit theorem \cite{Gnedenko1954}. 
If the microscopic dynamics of a many-body system exhibits these fluctuations, 
the aforementioned scheme of formulating the corresponding hydrodynamic EOM will no longer work 
since it is unclear how to perform an expansion in terms of the spatial derivatives of the hydrodynamic variables. 
Indeed, the difficulties arising in their formulation may explain why there is almost a complete lack of hydrodynamic theories for many-body systems with microscopic anomalous diffusive dynamics, with only some rare exceptions \cite{Grossman2016,Estrada2018}.

\begin{figure}[!tb]
	\centering
	\includegraphics[width=85mm,keepaspectratio]{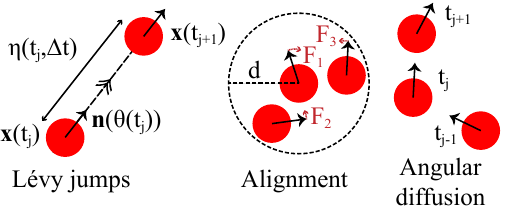}
	\caption{Schematic of the three processes captured in our microscopic model  with time discretization  
		$t_j\equiv j \Delta t$, where $\Delta t$ is a small time step.
		The random variable $\eta(t_j,\Delta t)$ represents the discretized step size, 
		which is sampled from a one-sided positive L\'evy stable distribution.
	}\label{figure1}
\end{figure}

In this Letter, we fill this knowledge gap 
by deriving hydrodynamic EOM 
for a two dimensional microscopic model of $N$ interacting active particles 
with power-law tailed distributed step sizes, 
which we call \textit{active L\'evy matter} (ALM).  
The dynamics of these active particles is akin to the L\'{e}vy flight model \cite{Mandelbrot1982,Hughes1981},
and is described by the Langevin equations 
(see schematic in Fig.~\ref{figure1})
	\begin{align}
	\dot{\vec{r}}_i(t)&=\eta_i(t) \vec{n}(\theta_i(t)) \ , & 
	\dot{\theta}_i(t)&=F_i(t) + \xi_i(t) 
	\label{model}
	\end{align}
where the index $i$ refers to the  $i$-th particle, 
the unit vector $\vec{n}(\theta_i)\equiv(\cos{\theta_i},\sin{\theta_i})$ prescribes its direction of motion,
and the  alignment force $F_i$ is given by
\begin{equation} 
F_i =
\left\{
\begin{array}{ll}
\frac{\gamma}{\pi d^2} \sum_{j=1}^N \sin{(\theta_j-\theta_i)}\ , &\  {\rm if\ } |\vec{r}_{ij}| \ ,
\\
0\ , &\ {\rm  otherwise} \ , 
\end{array}
\right.
\end{equation}
with $\vec{r}_{ij}\equiv \vec{r}_i-\vec{r}_j$ the distance between the $i$-th and $j$-th active particles and
$d>0$ the interaction range \cite{Vicsek1995,Peruani2008}. 
The angular noise $\xi$ is a white Gaussian noise of variance $\sigma$, i.e., 
$\langle \xi(t)\rangle=0$ and 
$\langle \xi(t)\xi(t^{\prime})\rangle=2\sigma\delta(t-t^{\prime})$ 
with $\langle \cdot \rangle$ denoting the averaging over $\xi$. 
The step size process $\eta$ is a one-sided positive L\'evy stable noise with characteristic function 
$\prec e^{i k \eta(t)} \succ=e^{-(-ik)^{\alpha}}$, where $0<\alpha<1$ and $\prec \cdot \succ$ denotes the averaging over $\eta$ \cite{Applebaum2009}.
The corresponding probability distribution has power-law tails $\propto \eta^{-(\alpha+1)}$ \cite{CairoliF}.
We note that for distributions with 
exponent $\alpha \geq 1$ 
the model is effectively equivalent to self-propelled dynamics with constant particle velocity \cite{Vicsek1995}.
We assume $\xi$ and $\eta$ to be independent  
and adopt the It\^o prescription in Eq.~\eqref{model}.

To derive 
hydrodynamic EOM for this model, 
we apply the BBGKY hierarchical formalism \cite{Huang1987} together with the simplest closure procedure to the hierarchical equations. 
Namely, we derive the Fokker-Planck equation for the single-particle distribution $P(\vec{r},\theta,t)$
by approximating the two-particle distribution $P_2$ as  
$P_2(\vec{r},\theta,\vec{r}^{\prime},\theta^{\prime},t)\simeq P(\vec{r},\theta,t)P(\vec{r}^{\prime},\theta^{\prime},t)$
\cite{CairoliF}.
This approximation is known to be 
suitable for dilute systems with long-range interactions such as plasmas \cite{Vlasov1968}.
Nevertheless, we believe that this also constitutes a good approximation in our model
because the L{\'e}vy step size statistics can effectively
render the binary interactions among the active particles long-ranged.
%
Using this closure and taking 
$d$ to be infinitesimal in the hydrodynamic limits,
we obtain the Fokker-Planck equation for the single-particle distribution
\cite{CairoliF}:
\begin{align}
\left(\derpar{}{t}+\mathcal{D}_{\vec{n}(\theta)}^{\alpha} - \sigma\dersecpar{}{\theta} - \derpar{}{\theta}M[P]\right) P&=0 \ , 
\label{1bodyeq}
\end{align}
where $M[P]$ is a functional of $P$ of the form
$ -\gamma\int_{-\pi}^{\pi} \sin{(\theta^{\prime}-\theta)} P(\vec{r},\theta^{\prime},t) \diff{\theta^{\prime}}$  
\cite{Bertin2006,
Baskaran2008,Peruani2008,Bertin2009,
Lee2010,Peshkov2014,Bertin2017}, 
and $\mathcal{D}_{\vec{n}(\theta)}^{\alpha} $ is the directional fractional derivative defined as \cite{Samko1993,Tarasov2011}
\begin{equation}
\mathcal{D}_{\vec{n}(\theta)}^{\alpha} P \equiv \frac{\alpha}{\Gamma(1-\alpha)}\int_0^{\infty} [1-\mathcal{T}^-_{\zeta\vec{n}(\theta)}]P \frac{\diff{\zeta}}{\zeta^{1+\alpha}} \ ,  
\label{fracder}
\end{equation} 
with the translation operator  
\beq
\mathcal{T}^{\pm}_{\zeta\vec{n}(\theta)}P(\vec{r},\theta,t)\equiv P(\vec{r}\pm\zeta\vec{n}(\theta),\theta,t)\ , 
\eeq
which is
$[-i (\vec{k}\cdot\vec{n}(\theta))]^{\alpha} \hat P(\vec{k},\theta,t)$
in spatial Fourier transform. 

With Eq.~\eqref{1bodyeq} at our disposal, we can construct systematically the hydrodynamic description of the microscopic model~\eqref{model}  in terms of 
the density field 
$\rho(\vec{r},t)\equiv \int_{-\pi}^{\pi} P(\vec{r},\theta,t) \diff{\theta}$,
the director field
$\vec{p}(\vec{r},t)\equiv \int_{-\pi}^{\pi} \vec{n}(\theta) P(\vec{r},\theta,t) \diff{\theta}$
and the nematic tensor 
$\boldsymbol{\mathrm{Q}}(\vec{r},t)\equiv \int_{-\pi}^{\pi} [\vec{n}(\theta)\vec{n}(\theta) - \frac{1}{2}\textbf{1} ]P(\vec{r},\theta,t) \diff{\theta}$, 
with $\textbf{1}$ being the $2\times 2$ identity matrix.
These fields are related to the  
lower order modes $f_0$, $f_1$ and $f_2$, respectively,
of the angular Fourier  transform of $P$ \cite{Bertin2006,
Baskaran2008,Peruani2008,
Bertin2009,Lee2010,Peshkov2014,Bertin2017}, which is defined as
$P(\vec{r},\theta,t)=(2\pi)^{-1}\sum_{m \in {\bf Z}} f_{m}(\vec{r},t)e^{-i m \theta}$,
where 
$f_m \equiv \int_{-\pi}^{\pi} e^{i m \theta} P(\vec{r},\theta,t)\diff{\theta}$. 
%
%
The temporal evolution equations of the modes $f_m$ can now be obtained from Eq.~\eqref{1bodyeq}; 
in spatially Fourier transformed space, these are given by 
\begin{multline}
\left( \derpar{}{t} + \Upsilon_0 k^{\alpha} + \sigma m^2 \right) \hat f_m  
+\frac{\gamma m}{2} (\hat f_{1+m} \star \hat f_{-1} - \hat f_{m-1} \star \hat f_{1}) \\ 
=-k^{\alpha}\sum_{m^{\prime}=1}^{\infty} i^{m^{\prime}} \Upsilon_{m^{\prime}} (e^{-i m^{\prime} \phi} \hat f_{m+m^{\prime}} + e^{i m^{\prime} \phi} \hat f_{m-m^{\prime}})  
\ ,
\label{amodes}
\end{multline}
where $\star$ denotes the convolution of two functions, i.e., 
$\hat f_i(\vec{k},t)\star \hat f_j(\vec{k},t)=\int_{-\infty}^{\infty} \diff{{}^2}{\vec{k^{\prime}}} \hat f_i(\vec{k}-\vec{k^{\prime}},t) \hat f_j(\vec{k^{\prime}},t)$,  
$\phi$ specifies the direction of the Fourier variable, i.e.,  
$\vec{k}\equiv k \, \vec{n}(\phi)$ with $k\equiv |\vec{k}|$ \cite{taylor2016}, 
and we have defined the following $\alpha$-dependent real and positive coefficients  \SM
\beq
\Upsilon_{m}(\alpha)\equiv \frac{(-i)^{m}}{2\pi} \int_{-\pi}^{\pi} e^{i \theta^{\prime} m}(-i \cos{\theta^{\prime}})^{\alpha}\diff{\theta^{\prime}}
\ .
\eeq

\begin{figure*}
\centering
\includegraphics[width=180mm,keepaspectratio]{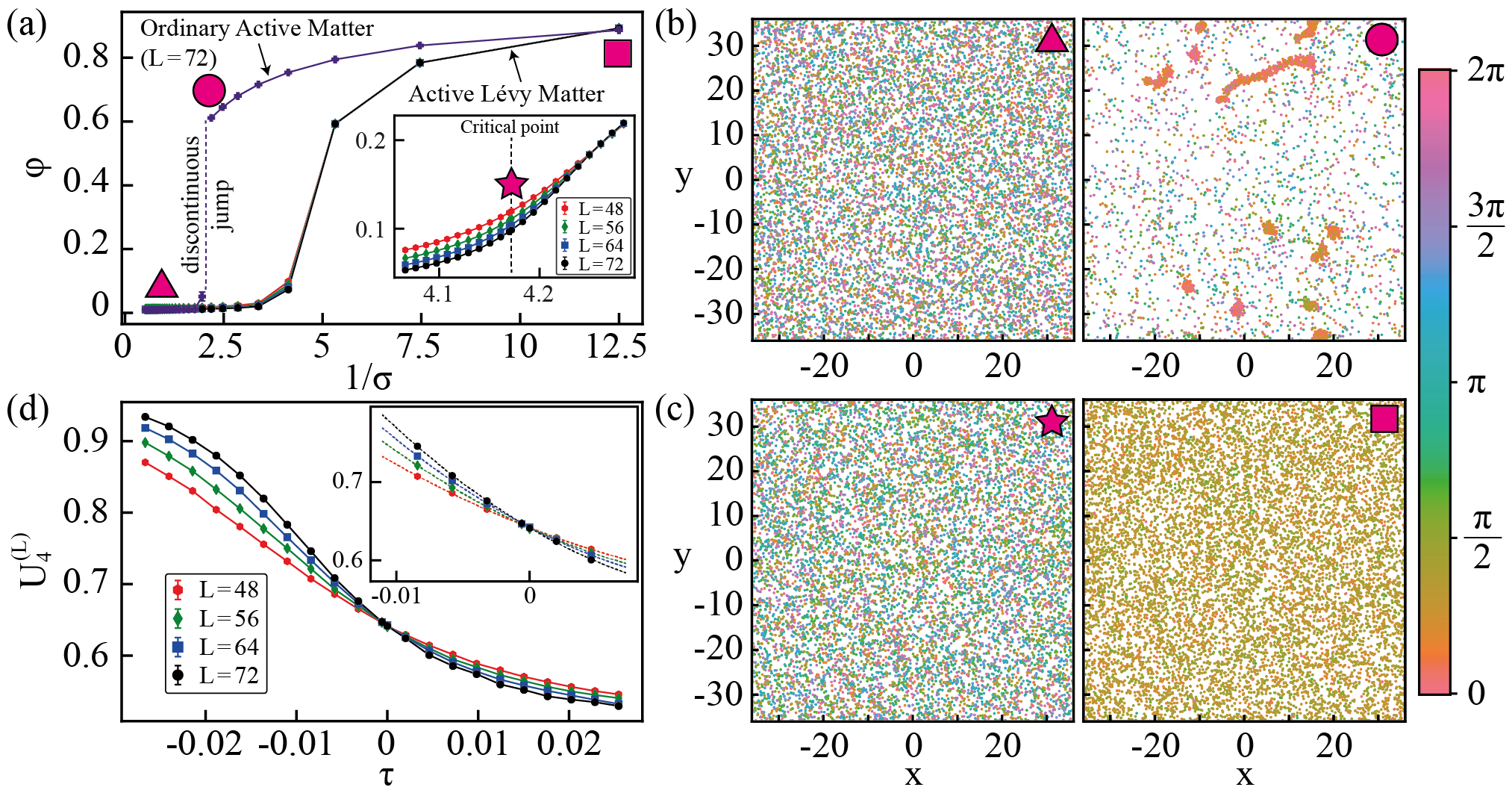}
\caption{
Finite size scaling analysis of the microscopic model of active L{\'e}vy matter. 
We run $300$ independent numerical simulations of the Langevin equations~\eqref{model} in a square box of linear length $L=\{ 48, 56, 64, 72\}$ with periodic boundary conditions.
Other parameters are $\rho=2$, $\gamma=0.25$ and $\alpha=1/2$ ($\alpha=1$ for Vicsek type dynamics). 
The model is initialized in the homogeneously disordered phase and updated for $2.5 \times 10^6$ time sweeps, which are sufficient for equilibration \SM.
(a) Time averaged polar order parameter $\varphi$ vs. the rotational noise strength $\sigma$.   
(b) Snapshots of the system configuration for ordinary active matter in the disordered phase (left) and in the unstable region at the onset of the transition (right). 
(c) Similar snapshots for active L\'evy matter at the critical point (left) and in the ordered phase (right). 
(d) Binder cumulant $U_4^{(L)}$ vs. the distance from the critical noise strength $\tau$. 
Inset: Quadratic fits of Binder cumulant curves around the critical point.  
In panels (a) and (d) error bars are plotted denoting 1 s.e.m.; in the inset of panel (d) we also plot 95\% confidence bands on the best fit lines. 
}\label{fig:simulations}
\end{figure*} 

Similar to ordinary active fluids, the hydrodynamic behaviour of the system around the onset of collective motion 
is dictated only by the lower order modes  \cite{Bertin2006,
Peruani2008,Bertin2009,Bertin2017,Lee2010,Peshkov2014},
because all the higher-order modes 
are suppressed by the ``mass'' term $\propto \sigma m^2$ 
and by the homogeneous term $\propto \Upsilon_0 k^{\alpha}$ resulting from the fractional operator~\eqref{fracder}.
%
We therefore adopt the same approximation strategy used in ordinary active matter, 
i.e., we assume that 
$\hat f_m\approx 0$ for $m\geq 3$ and 
$\partial_t \hat f_2\approx 0$.
The equation for $\hat f_2$ then becomes algebraic 
and can be solved. 
Using this exact result 
and re-expressing the equations for $f_0$ and $f_1$ in terms of the density, director, and nematic fields, 
we obtain the hydrodynamic EOM, to order $k^\alpha$, \SM
\begin{align}
&(\partial_{t} + \Upsilon_0 \mrD^{\alpha}) \rho=2\Upsilon_1 \mrI^{1-\alpha} (\nabla \cdot \vec{p}) 
- \lambda_2 \mrI^{2-\alpha} (\nabla\cdot\vecxi_0)  ,
\label{rhoeq} \\
&(\partial_{t} + \Upsilon_0 \mrD^{\alpha})\vec{p}=
[\kappa_0(\rho) - \xi |\vec{p}|^2 ]\vec{p}
+\Upsilon_1\mrI^{1-\alpha}\nabla\rho
\notag \\ & \quad
+ \lambda_1 \mrI^{1-\alpha}\vecxi_0
-\Upsilon_2\mrI^{2-\alpha}\nabla^2\vec{p}
+ \lambda_3 \mrI^{3-\alpha}\vecxi_1 
- \textbf{Q}\vec{p} \ ,
\label{meandireq}  
\end{align}
where
$\mrI^{\beta}$ is the fractional Riesz integral defined by its Fourier transform 
$\mrI^{\beta} h(\vec{r}) \rightarrow k^{-\beta} \hat h(\vec{k})$
(for $\Re{\beta}>0$ and $h$ a test function), 
and $\mrD^{\beta}$ is the corresponding fractional derivative 
whose Fourier representation is
$\mrD^{\beta} h(\vec{r}) \rightarrow k^{\beta} \hat h(\vec{k})$ \cite{Samko1993,Tarasov2011}. 
The operators $\mrI^{\beta}$ and $\mrD^{\beta}$ are the inverses of each other.  
In Eqs (\ref{rhoeq}) and (\ref{meandireq}), we have also defined the following
non-linear $(\vec{r},t)$-dependent vector fields
\begin{align}
\vecxi_0&\equiv 2[(\vec{p}\cdot \nabla)\vec{p}+(\nabla\cdot\vec{p})\vec{p}]-\nabla |\vec{p}|^2 \ , \label{xi0} \\
\vecxi_1&\equiv (\vec{p}\cdot\nabla)\nabla^2\vec{p}+(\vec{p}\times\nabla)\times\nabla^2\vec{p} \notag \\
&\quad +3[(\nabla\cdot\vec{p})\nabla^2\vec{p} + (\vecomega\times\nabla^2\vec{p}) - (\vecomega\cdot\nabla)\vecomega] \ , \label{xi1} 
\end{align}
where
$\vecomega(\vec{r},t)\equiv\curl{\vec{p}}$ 
is the vorticity field.
The nematic tensor $\textbf{Q}$ is specified by its components
$\text{Q}_{xx}=\mathcal{Q}_{1}=-\text{Q}_{yy}$ and $\text{Q}_{xy}=\mathcal{Q}_{2}=\text{Q}_{yx}$, 
which are 
\begin{align}
\mathcal{Q}_{i}
&=
-B_0\mrD^{\alpha}
\text{M}_{ij} p_j
+\frac{\lambda_1}{2} \mrI^{1-\alpha}
\text{T}_{ij}p_{j}
-\frac{\lambda_2}{4}\mrI^{2-\alpha}
\text{T}_{ij}\partial_j \rho
\notag \\
& \quad
+\frac{\lambda_3}{4}\mrI^{3-\alpha}
\text{T}_{ij}\nabla^2 p_j
-B_4 \mrI^{4-\alpha}
\text{T}_{ij}\Xi_{1 j} \ , 
\end{align}
where $i,j=1,2$ and $p_1$ and $p_2$ are the components of $\vec{p}$ along the $x$ and $y$ axis, respectively.
In the above, we have also defined the tensor function 
$\text{M}_{ij}\equiv p_{1} \delta_{ij} + p_{2} \epsilon_{ij}$ 
and the differential operator 
$\text{T}_{ij}\equiv \partial_{x} \delta_{ij} + \partial_{y} \epsilon_{ij}$, 
where $\delta_{ij}$ is the Kronecher symbol
and $\epsilon_{ij}$ is the generator of counter-clock wise rotations, 
with $\epsilon_{11}=0=\epsilon_{22}$ and $\epsilon_{21}=1=-\epsilon_{12}$.  
Furthermore, we have introduced the density-dependent quantity 
$\kappa_0(\rho)\equiv -\sigma+ (\gamma \rho)/2$ 
and the coefficients
\begin{align}
\lambda_1&\equiv \frac{\gamma \Upsilon_{1}}{4\sigma} \ , \ &
\lambda_2&\equiv \frac{\gamma \Upsilon_{2}}{2\sigma} \ , \ &
 \lambda_3&\equiv \frac{\gamma \Upsilon_{3}}{2\sigma} , \\
B_0&\equiv \frac{\gamma^2 \Upsilon_0}{32\sigma^2} \ ,\ &
B_4&\equiv \frac{\gamma^2 \Upsilon_4}{16\sigma^2} \ , \ &  
\xi&\equiv \frac{\gamma^2}{8\sigma} \ . 
\label{si-alphaP}
\end{align}

%
The hydrodynamic EOM (\ref{rhoeq}) and (\ref{meandireq}) are our first main result 
and may be viewed as 
the counterpart of the Toner-Tu equations for ALM. 
We note that, for $\alpha=1$, we have
$\Upsilon_i=0$ for all $i\neq1$ and 
$\Upsilon_1=-1/2$ {\SM}, so that
(\ref{rhoeq}) and (\ref{meandireq}) are reduced to 
a version of the Toner-Tu equations \cite{Toner1995,Toner1998,Toner2012} 
without the ordinary viscosity term, 
which is absent because of the truncation in the wavenumber $k$.

\begin{figure*}[!thb]
\centering
\includegraphics[width=180mm,keepaspectratio]{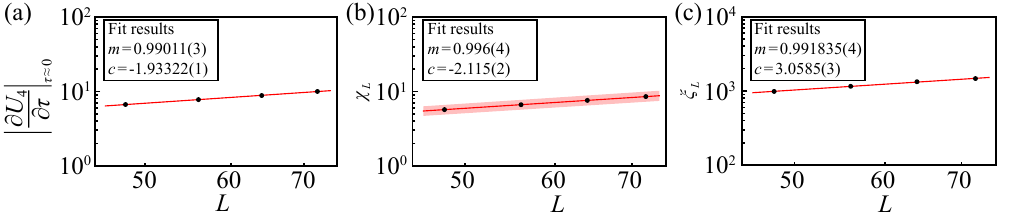}
\caption{Numerical estimation of static and dynamic critical exponents for the order-disorder transition in active L\'evy matter. 
Model parameters are $\rho=2$, $\gamma=0.25$ and $\alpha=1/2$. 
The simulation data (in logarithmic scales) are fitted with the linear function $m\,L+c$ by using a weighted least squares method. 
Error bars are plotted denoting 1 s.e.m.; 
95\% confidence bands on fit parameters are also shown. 
(a) Gradient of the Binder cumulant at criticality $|\partial U_4^{(L)}/\partial \tau|_{\tau \simeq 0}$ vs. the system size $L$. 
(b) Susceptibility at criticality $\chi_L$ vs. $L$. 
(c) Correlation time of the polar order parameter $\xi_L$ at criticality vs. $L$.}
\label{fig:crit_exps}
\end{figure*}

At the mean-field level, our EOM predict the same phase behaviour of the Toner-Tu equations, because
all terms involving (integral or fractional) spatial derivatives can be ignored. 
Equations~(\ref{rhoeq}) and (\ref{meandireq}) thus indicate 
(a) the disordered phase with $\vec{p}^*=0$ for $\sigma\geq\sigma_t$; and 
(b) the ordered phase with  
$\vec{p}^*\equiv\sqrt{8\sigma\kappa_0(\rho^*)/\gamma^2}\vec{e}_{\parallel}$, 
where $\vec{e}_{\parallel}$ denotes the arbitrary direction of spontaneous symmetry breaking 
for $\sigma<\sigma_t$. 
The density dependent threshold rotational noise strength is  
$\sigma_t(\rho^*)\equiv \gamma\rho^*/2$.
 
The mean-field description is known to be inadequate at the onset of collective motion for 
ordinary active matter due to the emergence of the characteristic banding instability,
which manifests itself as a longitudinal instability (i.e., along the direction of collective motion),
with the $k\rightarrow 0$ mode being the most unstable.
Strikingly, 
this is no longer true in ALM. 
To demonstrate this, we perturb the hydrodynamic fields 
$\rho$ and $\vec{p}$ as
$\rho^*+ \delta \rho_0 e^{s t +i \vec{q}\cdot\vec{r}}$ and 
$\vec{p}^*+\delta\vec{p}_0 e^{s t +i \vec{q}\cdot\vec{r}}$,
respectively, 
and solve (\ref{rhoeq}) and (\ref{meandireq}) at the linear level in $\delta \rho_0$ and $\delta\vec{p}_0$.
Eliminating them then
yields the dispersion relation with solutions $s_{\pm}(\vec{q})$ such that 
$\Re{s_+}>\Re{s_-}$. 
The spatially homogeneous phases predicted by the mean-field theory are only stable if $\Re{s_+}<0$.
Similar to ordinary active fluids, we find that
the disordered phase is stable against perturbations in all directions \SM. 
However, for the ordered phase, in the hydrodynamic limit, we obtain \SM
\begin{align}
\Re{(s_-)}&\simeq 
-2\kappa_0(\rho^*) + \mathcal{O}(q^{\alpha}) \ , \\ 
\Re{(s_+)}&\simeq
(-\Upsilon_0+2\Upsilon_2)q^{\alpha} + \mathcal{O}(q^{2\alpha}) \ .
\label{omegaP}
\end{align}
The first eigenvalue, $s_-$, always has a negative real part 
and describes the fast relaxation of small perturbations from $\vec{p}^*$. 
For normal active fluids, 
the real part of the second eigenvalue,  $s_+$, is always positive 
as the system approaches the onset of collective motion \cite{Bertin2006},  
indicating the banding instability  	
%
that renders the transition first-order \cite{Gregoire2004,Chate2008,Solon2013,Solon2015}.
For ALM, instead, this eigenvalue has a negative real part 
because $\Upsilon_0>2\Upsilon_2$ for all $0<\alpha<1$. 
Therefore, the banding instability is made stable by the L\'{e}vy motion of the active particles.
Furthermore, we can demonstrate that transversal perturbations (i.e., orthogonal to the direction of collective motion) 
are also suppressed \cite{CairoliF}.
As a result, in ALM, the ordered phase always remains stable at the onset of collective motion in the hydrodynamic limit, thus rendering the order-disorder phase transition potentially critical.
These predictions represent the second main result of our work. 

%

\begin{table}[!b]
\setlength{\tabcolsep}{0.75em}
\begin{tabular}{ccccc} 
$\nu$ & $\beta/\nu$ & $\gamma/\nu$ & $z$ \\ 
\hline\hline
& & & \\ [\dimexpr-\normalbaselineskip+2pt]
	$1.00999(3)$ 
& 	$0.497(1)$ 
& 	$0.996(4)$ 
& 	$0.991835(4)$ \\ 
& & & \\ [\dimexpr-\normalbaselineskip+2pt]
\hline
\end{tabular}
\caption{Numerical estimates of the critical exponents for $\alpha=1/2$. Errors are expressed as 1 s.e.m.. 
}\label{tab:critical}
\end{table}

We now investigate the critical behavior predicted by our theory with numerical simulations. We focus here on the particular value $\alpha=1/2$ \cite{CairoliF}.
We first note that 
the criticality of the phase transition manifests in
the smooth approach of the time averaged polar order parameter $\varphi$ 
to zero as the rotational noise is increased, 
as opposed to the abrupt jump found in ordinary active matter, 
which is characteristic of a first-order transition (Fig.~\ref{fig:simulations}a). 
Snapshots of the system configuration 
also confirm these predictions
(Figs.~\ref{fig:simulations}b and \ref{fig:simulations}c).
We now characterize the critical exponents of the phase transition
by performing a finite size scaling analysis \cite{Landau2014}. 
Specifically, we first estimate the asymptotic critical noise strength at fixed density, 
$\sigma_{\star}$,
by identifying the crossing of the Binder cumulant 
$U_4^{(L)}\equiv \langle \varphi^2 \rangle_{L}^2/ \langle \varphi^4 \rangle_{L}$
for different system sizes $L$ (Fig.~\ref{fig:simulations}d).
Here, $\langle \cdot \rangle_L$ denotes the ensemble average taken at finite system size $L$.
We estimate: 
$\sigma_{\star}=0.239816(2)$ (error is 1 s.e.m.).
Using the scaling relation
$|\partial U_4^{(L)}/\partial \tau|_{\tau \simeq 0}\propto L^{1/\nu}$, 
where $\tau\equiv -1 + \sigma/\sigma_{\star}$, 
we can estimate directly the static exponent $\nu$ (Fig.~\ref{fig:crit_exps}a).   
In addition, using the scaling relations
for the polar order parameter at criticality  
$\varphi_{\star} \propto L^{-\beta/\nu}$
and that for the susceptibility
$\chi_L \propto L^{\gamma/\nu}$, 
with 
$\chi_L \equiv L^2 (\langle \varphi_{\star}^2 \rangle_L - \langle \varphi_{\star} \rangle_L^2)$,
we can estimate the ratios of static critical exponents $\beta/\nu$ and $\gamma/\nu$ (Fig.~\ref{fig:crit_exps}b).
To estimate the dynamic exponent $z$, we use the scaling relation for the correlation time of the polar order parameter 
$\xi_L \propto L^z$ (Fig.~\ref{fig:crit_exps}c).
The estimates obtained are summarized in Table~\ref{tab:critical}. 
The numerical characterization of the critical properties of the order-disorder transition is our third main result.
Intriguingly, these estimates for $\alpha=1/2$ seem to indicate that the static properties of the critical transition in ALM (for the particular $\alpha$ considered here) belongs to the same universality class of the equilibrium model in two spatial dimensions with long-range interaction energy $\propto 1/|\vec{r}_{ij}|^{2+2\alpha}$ 
and $n$-component order parameter
\cite{Fisher1972,Suzuki1973}.
Elucidating analytically this connection by employing dynamic renormalization group methods, 
similarly to what was accomplished for ordinary active matter in incompressible conditions \cite{Chen2015,Chen2016,Chen2018}, 
is an interesting open problem that 
we aim to elucidate in future investigations.



In this Letter, we derived the first hydrodynamic description of ALM, 
which generalizes the conventional theory of active fluids 
by incorporating L\'evy stable distributed fluctuations with diverging variance.
We then revealed that,
unlike ordinary polar active matter, 
the order-disorder transition in ALM 
is critical 
and estimated the corresponding critical exponents numerically. 
Our work highlights the novel physics exhibited by active matter models integrating both anomalous diffusive motility and inter-particle interactions.  
Interesting future directions include the investigation of the effects of L\'{e}vy displacement dynamics on 
collective phenomena other than collective motion 
such as active turbulence 
\cite{Dombrowski2004,Hernandez-Ortiz2005,Sokolov2007,Aranson2007,Saintillan2007,Wolgemuth2008,
Sanchez2012,Wensink2012,Doostmohammadi2018}
and motility-induced phase separation
\cite{Tailleur2008,Fily2012,Redner2013,Cates2015}, 
as well as their relevance to biological systems \cite{Zaburdaev2015}. 




\begin{acknowledgments}
	A.~C. gratefully acknowledges funding under the Science Research Fellowship granted by the Royal Commission for the Exhibition of 1851,  
	and the High Throughput Computing service provided by Imperial College Research Computing Service, DOI: 10.14469/hpc/2232. 
\end{acknowledgments}


%

\end{document}